\documentclass[12pt]{article}
\usepackage{epsfig,wrapfig}
\usepackage{amssymb}
\def\eq#1{\begin{equation} #1 \end{equation}}

\let\la=\lesssim            
\let\ga=\gtrsim

\begin{document}

\begin{center}{\Large {\bf Radiation pressure mixing of large dust grains in
protoplanetary disks }}\\
{\bf Dejan Vinkovi\'{c}}\\
Physics Department, University of Split, \\
Nikole Tesle 12, 21000 Split, Croatia\\
e-mail: vinkovic@pmfst.hr\\
{\bf For the final print version see: Nature 459, 227-229 (14 May 2009)}
\end{center}


{\bf

Dusty disks around young stars are formed out of interstellar dust that consists of amorphous, submicrometre grains. Yet the grains found in comets \cite{Brownlee} and meteorites \cite{Wooden_05}, and traced in the spectra of young stars \cite{Boekel_04}, include large crystalline grains that must have undergone annealing or condensation at temperatures in excess of 1,000 K, even though they are mixed with surrounding material that never experienced temperatures as high as that \cite{Hill_01}. This prompted theories of large-scale mixing capable of transporting thermally altered grains from the inner, hot part of accretion disks to outer, colder disk regions \cite{Xwind,Ciesla_07,Boss08}, but all have assumptions that may be problematic \cite{Klahr_06, Hubickyj_06, Matsumura_06,Shang_07,
Boss05}. Here I report that infrared radiation arising from the dusty disk can loft grains bigger than one micrometre out of the inner disk, whereupon they are pushed outwards by stellar radiation pressure while gliding above the disk. Grains re-enter the disk at radii where it is too cold to produce sufficient infrared radiation pressure support for a given grain size and solid density. Properties of the observed disks suggest that this process might be active in almost all young stellar objects and young brown dwarfs.}


The history of thermal and compositional alternation of dust in
dense dusty protoplanetary disks around young pre-main-sequence
(PMS) stars enables us to better understand conditions that initiate
formation of planets. One of the long standing problems arising from
this approach is the presence of crystalline dust in disk
environments considered too cold for crystallinity to occur. Thus,
it has been suggested that silicates crystalize in the hot part of
the disk close to the central star and are transported outward into
colder environment. Currently favored theories of outward transport
include: i) turbulent mixing \cite{Ciesla_07}, ii) ballistic
launching of particles in a dense wind created by interaction of the
accretion disk with the young star's magnetic field (X-wind
model)\cite{Xwind}, and iii) mixing mediated by transient spiral
arms in marginally gravitationally unstable disks\cite{Boss08}.
Although these theories sound promising and may eventually result in
the definitive solution to the problem of a large scale mixing, they
are so far hampered by theoretical assumptions needed for them to
work. The turbulent mixing requires a source of efficient turbulent
viscosity and the magnetorotational instability (MRI) is invoked as
the most promising candidate, but large stretches of the disk are
considered not sufficiently ionized to keep MRI active
\cite{Klahr_06, Hubickyj_06, Matsumura_06}. The X-wind model relies
on the theoretical notion of magnetic field configurations in the
immediate vicinity of PMS stars and high hopes are put on future
observations to resolve this predicament \cite{Shang_07}. The spiral
arms model is in the domain of discussions whether the underlying
numerics, physical approximations and assumptions on the initial
conditions are realistic enough to make results plausible
\cite{Klahr_06, Hubickyj_06, Boss05}.

Unlike these theories, non-radial radiation pressure does not
require additional assumptions on the physical conditions in the
disk because it stems from the basic radiative transfer properties
of optically thick dusty disks. It has been already shown that
individual submicron grains do not move far away in the disk when
pushed by radiation pressure because the force is primarily produced
by radial stellar flux \cite{Takeuchi_03}. On the other hand, micron
or larger grains are large enough to also have efficient interaction
with the near infrared photons (NIR; equivalent to dust temperatures
of $\sim$1000-2000K) from the hot inner disk. Submicron grains are
very inefficient emitters in NIR, hence, they overheat and sublimate
further away from the inner disk surface. This leaves the surface
populated only with large grains, while small grains can survive
within the optically thick interior \cite{Vinkovic_06} or at larger
disk radii. Direct imaging with NIR interferometers revealed that
the observed location of inner disk rim is consistent with this
description (see \cite{Millan-Gabet_07,Vinkovic_07,Isella_05} and
references therein).

In optically thick protoplanetary disks dust particles $\la 1mm$ are
well coupled with the gas and their dynamics is dominated by the gas
drag \cite{Takeuchi_03,Alex08}. Hence, dust motion is very similar to the gas
orbital, almost Keplerian, motion. Radiation pressure force serves
as a slow perturbation that leads to the rearrangement of dust
orbits. In order to reconstruct the trajectory of particles pushed
by radiation we need to derive the spatial orientation of radiation
pressure vector. For that, we need estimates of the diffuse flux as
the source of pressure asymmetry. I solve this using the two-layer
formalism, which is a well established method utilized in problems
involving protoplanetary disk emission \cite{Chiang_97}.

\begin{figure}[t]
 \includegraphics[width=6in]{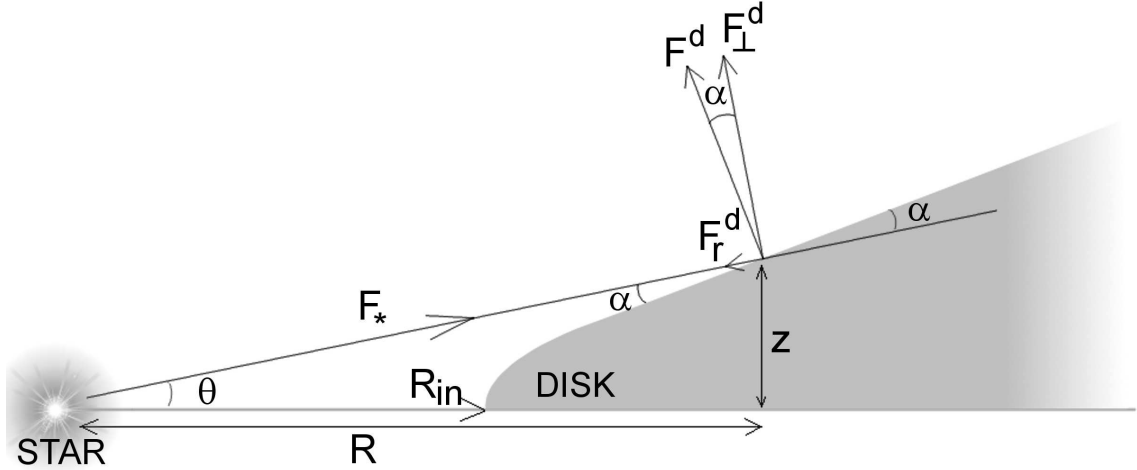}
 \caption{\small {\bf Geometry of non-radial radiation pressure.}
The sketch shows a cross section of an optically thick protoplanetary
dusty disk heated by a star. The disk has a central hole of radius
$R_{in}$ where the dust overheats and sublimates away. According to
the two-layer formalism \cite{Chiang_97}, analysis of the disk
emission in the near IR can be reduced to the disk's optically thin
surface, which is heated directly by the stellar radiation. In this
approach the disk surface is replaced with a single temperature
layer and we assume that the stellar radiation is completely
absorbed within this layer. The disk interior is described as the
second temperature layer, but it is heated only by infrared
radiation from the surface layer and, therefore, it is much colder
and does not contribute to the disk emission in the near and mid IR
\cite{Chiang_97}. In optically thick passive disks we can use energy
conservation at a surface point $(R,z)$ to set balance, $F_*
\sin\alpha=F^d$, between the bolometric stellar flux $F_*$
intercepted by the disk at a grazing angle $\alpha$ and the outgoing
disk radiation $F^d$ (IR emission and scattered stellar photons).
In the approximation of geometrically thin disk surface we can assume
that the entire local diffuse flux at the very surface is perpendicular
to the surface. Grains that manage to decuple and move away from the surface
would feel a reduced flux since the diffuse radiation streams out in all directions.
We can decompose $F^d$ into radial, $F^d_r=-F_*
\sin^2\alpha$, and azimuthal, $F^d_\bot=F_* \sin\alpha\cos\alpha$,
components. If dust grains are big enough to have constant
extinction in the wavelength range of $F^d_\lambda$ then the
radiation pressure force becomes $\vec{\mathfrak{F}}\propto
\vec{F_*}+\vec{F_d}$. Using flux components from above gives the
radial force $\mathfrak{F}_r\propto F_*\cos^2\alpha$ and the
azimuthal force $\mathfrak{F}_\bot\propto F_*\sin\alpha\cos\alpha$.
Notice that this yields radiation pressure force directed exactly
parallel to the disk surface,
$\mathfrak{F}_\bot/\mathfrak{F}_r=\tan\alpha$, irrespective of the
disk curvature. A more rigorous derivation is presented in
Supplementary information \S1, including dust dynamics due to
gravity, gas drag and radiation pressure.
 }
 \label{Figure1}
\end{figure}

A short simplified solution is presented in figure \ref{Figure1},
while a more rigorous derivation, which includes gravity, gas drag
and radiation pressure, is described in Supplementary information
\S1. The result shows that the net radiation pressure force, which
combines stellar and diffuse flux components, is directed exactly
parallel to the disk surface irrespective of its curvature.
This leads to a very interesting scenario. If the force is strong
enough to move a large dust grain then such a large crystalline grain
formed at the hot inner rim would glide over the disk surface toward
colder disk regions until the diffuse disk flux becomes too ``cold''
(i.e. its peak wavelength is larger than the dust size), at which
point the force keeping the dust afloat ceases.

\begin{figure}[t]
 \includegraphics[width=6in]{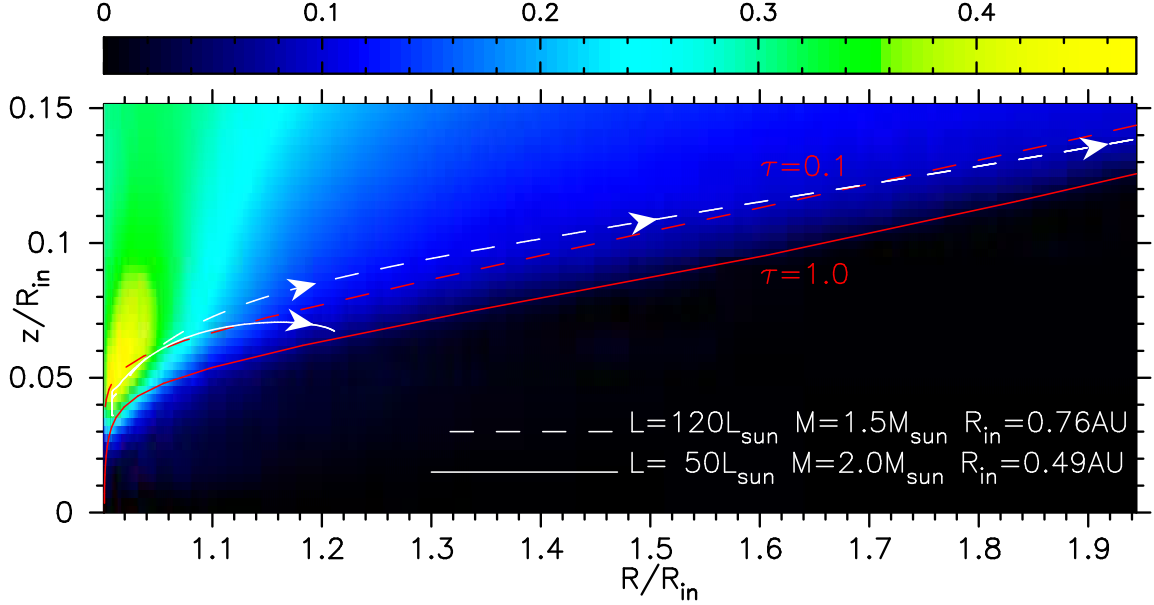}
 \caption{\small {\bf Trajectory of dust grains under the influence of stellar gravity,
gas drag and non-radial radiation pressure.}
The colored background map shows the vertical $z$ component of the radiation pressure
vector scaled with the value that the stellar pressure would have if
the dusty disk were not there. Two stellar luminosty-mass ratios are
used: 80$L_\odot/M_\odot$ (white dashed line) and
25$L_\odot/M_\odot$ (white solid line). Dust composition is
olivine \cite{Dorschner95} of 2$\mu$m radius and solid density
3000$kg/m^3$. Dust grains start their travel with a vertical upward
motion until the gas density drops enough to loosen the influence of
gas drag. After that the grain is ejected to a larger disk radius,
where trajectory details depend on the strength and direction of
radiation pressure. Trajectory is calculated numerically with the
Runge-Kutta method. Radiation pressure is calculated numerically
from 2D radiative transfer that includes dust absorption, scattering
and emission. The disk consists of 0.1$\mu$m and 2$\mu$m grains that
sublimate at 1500K, but the surface in this disk region is too hot
for 0.1$\mu$m grains, which survive below the surface populated by
2$\mu$m grains. Spatial dimensions are scaled with the disk
sublimation radius $R_{in}$ (see figure \ref{Figure1}). The disk gas
and dust densities decrease exponentially with $z$. Red lines show
the disk surfaces defined by the radial visual optical depth of 0.1
(dashed red line) and 1 (solid red line). Details of
this numerical result will be shown in a separate publication. }
 \label{Figure3}
 \end{figure}

Further insight into the nature of non-radial radiation pressure
outflow requires a more detailed description of the disk structure
and an advanced radiative transfer calculation. I started with
preliminary modeling at such an advanced level. The first results
are presented in figure \ref{Figure3}. The model assumes dusty disk
density structure of the form $\rho_d(R,z)\sim
R^{-2}\exp(-z^2/2h^2)$, with the scale height $h=1.67\times
10^{-2}R^{1.25}$, where $ R$ and $z$ are cylindrical coordinates
scaled with the dust sublimation radius  $R_{in}$ (the disk's inner
rim; see figure \ref{Figure1}). The disk contains 0.1$\mu$m and
2$\mu$m olivine grains with the relative density ratio 10$^4$:1 and
the overall radial visual optical depth at $z=0$ of 10,000. I
performed a full 2D radiative transfer for the case of disk heating
from a 10,000K star and 1,500K dust sublimation. Location of the
dust sublimation disk surface is calculated self-consistently from
the mutual exchange of infrared energy between 0.1$\mu$m and 2$\mu$m
grains, resulting in $R_{in}=44.7 R_*$ ($R_*$ is the stellar
radius). Figure \ref{Figure3} shows the map of vertical radiation
pressure on 2$\mu$m grains and examples of grain trajectories.
Results from this detailed approach confirm the plausibility of our
theoretical arguments.

The ability of large grains to migrate along any disk curvature makes
this theory independent of the ongoing debate on the geometrical
structure of the inner disk region \cite{Millan-Gabet_07}. The
popular view is that the inner sublimation edge is puffed up and
curved \cite{Isella_05}. The non-radial pressure would affect dust
dynamics under such a disk curvature in the same way as in the
numerical example above, except that individual grains could
decouple more easily from the inner disk and fly toward outer disk
regions due to the disk's self-shadowing \cite{Fujiwara_05}.

Grains pushed by radiation create an outflow that operates at much
shorter timescale than the local dust settling because radiation
pressure is active in the region of lower gas density. In
Supplementary information \S2 I provide an estimate of the total
amount of dust that flows outward in the disk surface layer. The
outflow strength and range depend on the ratio, $\beta$, of
radiation force tangentially to the disk surface over local
gravity force (see Supplementary information \S1 for a detailed
description):
 \eq{\label{beta}
  \beta\sim 0.4\,\,\left[{L_*\over
  L_\odot}\right]\,\,\left[{M_\odot\over M_*}\right]\,\,\left[{3000
  kg/m^3\over\varrho_s}\right]\,\,\left[{\mu m \over a}\right],
 }
where $a$ is the dust grain radius, $L_*$ is the stellar luminosity,
$M_*$ is the stellar mass and $\varrho_s$ is the grain solid
density. Grains with $\beta \ga 0.5$ are gravitationally decoupled
from the star and will be pushed away from the star as long as the
diffuse flux keeps them afloat within the optically thin surface.
Grains with $\beta \la 0.5$ feel a ``reduced'' gravity and their
settling is slowed down.

\begin{figure}[t]
 \includegraphics[width=5in]{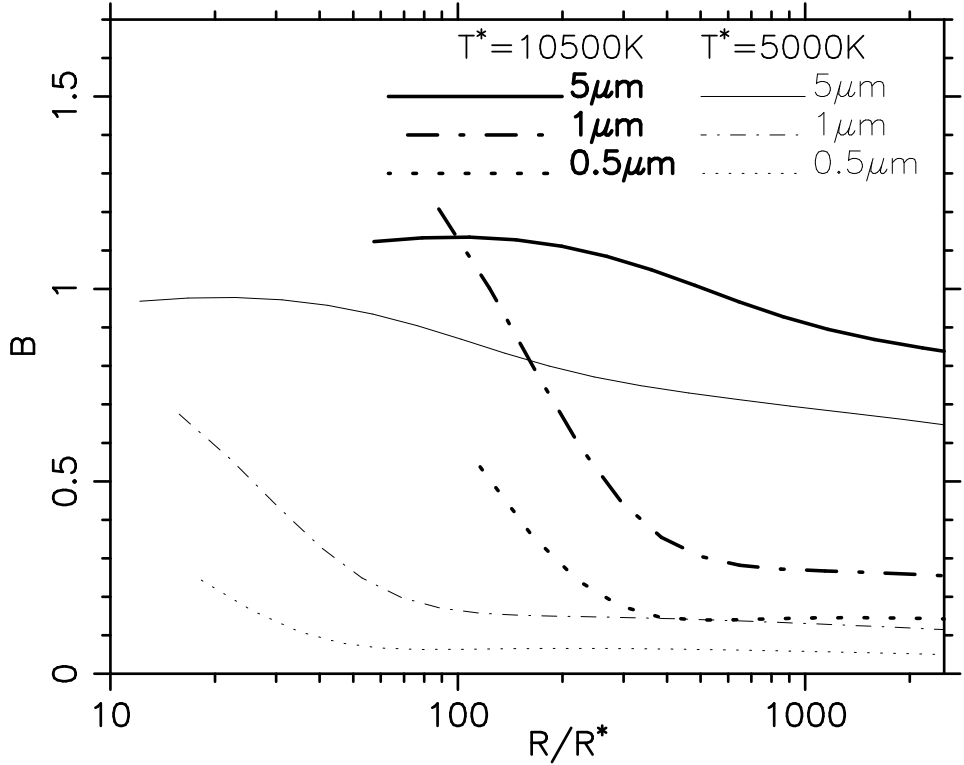}
 \caption{\small {\bf Estimated strength of diffuse radiation pressure along the
disk surface, indicating how far grains can travel.}
 The estimated strength, $B$, is defined
as the ratio of diffuse to stellar $\beta$ perpendicular to the disk
surface (see equation 34 in Supplementary information
\S3 for details), at various distances from the star. Optical
properties of the pushed grains and dust forming the disk surface
are the same. The surface contains only one grain size and type.
Lines show results for spherical grains of 0.5$\mu$m, 1$\mu$m and
5$\mu$m radius. Diffuse radiation pressure is important (i.e. $B\sim
1$) only for grains $\ga$1$\mu$m. Grains $\ga$5$\mu$m experience
strong diffuse pressure over a large disk surface because of their
efficient infrared absorption at longer wavelengths, while smaller
micron grains can float only above the inner disk with the highest
temperature. The dust is enstatite from \cite{Jaeger98}. Other
compositions lead to qualitatively similar curves. Lines start at
radii defined by 1,500K dust sublimation temperature.}
 \label{Figure2}
 \end{figure}

I made an attempt to estimate the spatial extent of significant
vertical radiation pressure along the disk surface. I use a
simplified, but illustrative model of the protoplanetary disk where
the disk surface contains only single size grains. Results show (see
figure \ref{Figure2} and Supplementary information \S3) that
significant dynamical effects from the non-radial radiation pressure
are possible only for grains larger than about 1$\mu$m. Grains a few
microns in size can be lifted out of the disk only at small disk
radii where the disk is the hottest, but already 5$\mu$m grains can
``glide'' to large radii (over 1,000 stellar radii), provided that
the radiation pressure is strong enough to push such a grain. The
upper limit on grain size pushed that way is dictated by equation
\ref{beta} that shows how the force decreases with grain size.

Notice that I assume a solid spherical grain, which is a
simplification of a more realistic fluffy dust aggregate
\cite{Voshchinnikov08,Pinte08}. Aggregates result in a much larger
$\beta$ for the same grain size because they have a much lower grain
density than the typical $3000 kg/m^3$ due to inclusion of vacuum
into the grain structure. On the other hand, crystalline grains are
largely transparent in the spectral range of stellar radiation
\cite{Brownlee}, which would make radiation pressure ineffective.
This remains an open problem for our theory, although crystalline
grains incorporated into dust aggregates might have a
non-transparent ``glue" keeping the aggregate together, which would
increase $\beta$ and mitigate these problems. Such ``dustballs" are
considered to be precursors of chondrules and CAIs in meteorites
\cite{Jones_00}.

The main stellar parameter dictating the overall strength $\beta$ of
radiation pressure effect on a grain is the luminosity-mass ratio
$L_*/M_*$. Observations and evolutionary tracks indicate that
\hbox{$L_*/M_*\ga 0.5$} (which gives $\beta\sim 0.4$ for a grain of
1$\mu$m diameter) in almost all young stellar objects, including
brown dwarfs. Thus, non-radial radiation pressure is at least
marginally relevant in all these objects, especially if a realistic
dust aggregate model is taken into account. Moreover, at earlier
evolutionary stages $\beta$ was larger because, according to stellar
evolution models, the end of significant accretion (99\% of the
final mass) ends with $L_*/M_* > 10$ for stars $M_*\la 1M_\odot$
\cite{Wuchterl03}.

Since crystallization is very efficient along the hot inner disk
rim, radiation pressure mixing of large grains would inevitably
include the crystalline fraction and disperse such dust over the
disk surface. Interestingly enough, such a correlation between large
grains and crystalline fraction is detected in Herbig Ae stars
(e.g.\cite{Boekel_05,Sargent_06}). This would be the most pronounced
in the inner disk regions, closer to the inner rim, as it is indeed
observed (e.g. \cite{Boekel_04,Watson_07,Schegerer_08}). With the
help from disk turbulence, the surface of inner disk region is
constantly replenished with new grains and the process continues as
long as the radiation pressure is active.

\noindent
{\bf Supplementary Information} is linked to the online version of the
paper at www.nature.com/nature.

\noindent {\bf Acknowledgements}: The author thanks the Institute
for Advanced Study in Princeton and the University Computing Center
SRCE in Zagreb for time on their computer clusters.

\noindent {\bf Author Information}: Reprints and permissions
information is available at\\ npg.nature.com/reprintsandpermissions.
The author declares no competing financial interests. Correspondence
and requests for materials should be addressed to vinkovic@pmfst.hr.

\newpage

\newpage

\newpage

\newpage
\onecolumn

\renewcommand{\thepage}{Supplementary information: page \arabic{page}}
\setcounter{page}{1}

 \setcounter{section}{0}

\begin{center}
{\bf \Large SUPPLEMENTARY INFORMATION}
\end{center}

\newpage

\section{{\Large Equations for non-radial radiation pressure
dynamics}}

The radiation pressure force pushing a grain of radius $a$ in
direction $\hat{n}$ is
 \eq{
  \overrightarrow{\mathfrak{F}}={a^2\pi\over c}\int Q^{ext}_\lambda
  \overrightarrow{F_\lambda}d\lambda,
 }
where $c$ is the speed of light, $Q^{ext}_\lambda$ is the extinction
coefficient and $\overrightarrow{F_\lambda}$ is the total radiation
flux in direction $\hat{n}$. The stellar contribution to the
radiation pressure on a grain at distance $r$ from the star is
 \eq{
   \mathfrak{F}_*={a^2\pi\over c}\int Q^{ext}_\lambda
  F_{*\lambda}d\lambda={L_*a^2\over 4cr^2}\int
  Q^{ext}_\lambda f_{*\lambda}d\lambda,
 }
where $L_*$ is the stellar luminosity and $f_{*\lambda}$ is
normalized shape of stellar spectrum $\int f_{*\lambda}d\lambda=1$.
We use $\mathfrak{F}_*$ to scale the force
 \eq{
 {\overrightarrow{\mathfrak{F}}\over \mathfrak{F}_*}= \overrightarrow{\mathfrak{f}} =
 {\int Q^{ext}_\lambda
  \overrightarrow{F_\lambda}d\lambda\over \int Q^{ext}_\lambda
  F_{*\lambda}d\lambda}=
  \hat{r}+{\int Q^{ext}_\lambda
  \overrightarrow{F^d_\lambda}d\lambda\over \int Q^{ext}_\lambda
  F_{*\lambda}d\lambda}.
 }
In the case of a negligible diffuse flux the pressure becomes
identical to the stellar radiation force and
$\overrightarrow{\mathfrak{f}}=\hat{r}$.

If dust grains are big enough to have a constant extinction in the
wavelength range of $F^d_\lambda$ then the radiation pressure force
becomes $\vec{\mathfrak{F}}\propto \vec{F_*}+\vec{F_d}$. The disk
flux is perpendicular to the disk surface and we can decompose it
into radial and azimuthal components (see figure \ref{Figure1})
 \eq{\label{flux_d_r}
   F^d_r=-F_* \sin^2\alpha,
 }
 \eq{\label{flux_d_t}
   F^d_\bot=F_* \sin\alpha\cos\alpha.
 }
Using these components gives the radial force
 \eq{\label{force_r}
   \mathfrak{F}_r\propto F_*-F_*\sin^2\alpha = F_*\cos^2\alpha
 }
and the azimuthal force
 \eq{\label{force_p}
  \mathfrak{F}_\bot\propto F_*\sin\alpha\cos\alpha.
  }

We see from this analysis that the radiation pressure force is
directed exactly parallel to the disk surface
$\mathfrak{F}_\bot/\mathfrak{F}_r=\tan\alpha$ {\it irrespective of
the disk curvature}. This leads to a very interesting scenario. If
the force is strong enough to move a big dust grain then such a big
crystalline grain formed at the hot inner rim would glide over the
disk surface toward colder disk regions until the diffuse disk flux
becomes to ``cold'' (i.e. its peak wavelength is larger than the
dust size), at which point the force keeping the dust afloat ceases.

The ``strength'' of the radiation pressure force is measured by the
ratio of the radiation pressure in direction $\hat{n}$ and the local
gravity force
 \eq{
  \overrightarrow{\beta}={ \overrightarrow{\mathfrak{F}}\over
  GM_*m_d/r^2} = {L_* a^2\over 4cGM_*m_d}\overrightarrow{\mathfrak{f}}\int
  Q^{ext}_\lambda f_{*\lambda}d\lambda,
 }
which yields
 \eq{\label{eq_beta}
  \overrightarrow{\beta}=0.2\,\,\left[{L_*\over
  L_\odot}\right]\,\,\left[{M_\odot\over M_*}\right]\,\,\left[{3000
  kg/m^3\over\varrho_s}\right]\,\,\left[{\mu m \over a}\right]\overrightarrow{\mathfrak{f}}\int
  Q^{ext}_\lambda f_{*\lambda}d\lambda,
 }
where $\varrho_s$ is the solid density of a grain and $L_\odot$ and
$M_\odot$ are solar luminosity and mass. Grains with
$\beta\cos\alpha \ge 0.5$ are gravitationally decoupled from the
star and will be pushed away from the star as long as the diffuse
flux keeps them afloat within the optically thin surface. Grains
with $\beta\cos\alpha < 0.5$ ``feel'' a reduced gravitational force
and shift to a larger stable orbit.

The pressure vector $\overrightarrow{\mathfrak{f}}$ is close to
unity and, as described above, points tangentially to the disk
surface. Stellar radiation peaks at wavelengths smaller than big
grains, hence we can approximate $Q^{ext}_\lambda\sim2$ and $\int
Q^{ext}_\lambda f_{*\lambda}d\lambda\sim 2$. This value of
$Q^{ext}_\lambda$ is correct independently of all other grain
properties (chemical composition or shape) when the grain size is
much larger than the wavelength (so called ``extinction paradox'').
In cases when the grain size is larger by a factor of a few,
$Q^{ext}_\lambda$ can vary between $\sim 1$ and $\sim 4$ due to the
contribution of anisotropic scattering. We see from this that the
stellar luminosity to mass ratio and grain density dictate the size
of a grain capable of migrating out of the hot inner disk rim.

The equation of motion of a particle in a gaseous medium includes
forces of gravity, gas drag in the Epstein regime and radiation
pressure \cite{Takeuchi_03,Garaud04}
 \eq{
  \ddot{\overrightarrow{r}} = -G{M_*\over r^3}\overrightarrow{r}-
  {\varrho_g\over \varrho_s}\,\,{c_s\over
  a}(\dot{\overrightarrow{r}}-\overrightarrow{v_g}) +
  \overrightarrow{\beta}G{M_*\over r^2},
 }
where $\varrho_g$ and $\overrightarrow{v_g}$ are the local gas
density and velocity, respectively, and $c_s$ is the local speed of
sound. We can assume that on the time scales of interest the gas has
no radial or vertical velocities and rotates around $\hat{z}$ axis
with the Keplerian angular velocity. We expand this equation in a
cylindrical coordinate system $( R,\varphi,z)$
 \eq{\label{eq_rho}
 \ddot{ R}= R \dot{\varphi}^2 - {GM_*\,\, R\over
 ( R^2+z^2)^{3/2}}- \mu\dot{ R} + \beta_ R{GM_*\over
  R^2+z^2},
 }
 \eq{\label{eq_phi}
 \ddot{\varphi}=-2\,{\dot{ R}\over R}\,\dot{\varphi} -
 \mu\left(\dot{\varphi}-{v_g\over R}\right),
 }
 \eq{\label{eq_z}
 \ddot{z}= - {GM_*\,\,z\over
 ( R^2+z^2)^{3/2}}- \mu\dot{z} + \beta_z{GM_*\over  R^2+z^2},
 }
where we use $\overrightarrow{\beta}=\beta_ R\hat{
R}+\beta_z\hat{z}$ and
 \eq{\label{eq_mu}
  \mu={\varrho_g\over \varrho_s}\,\,{c_s\over a}.
 }
We work in the regime $\mu\gg \dot{ R}/ R$ where particles are
strongly coupled with the gas and have a short gas drag stopping
time. Hence, in equation \ref{eq_phi} we can assume that dust and
gas have the same angular motion similar to the Keplerian speed
$\Omega^2_K=GM_*/ R^3$ \cite{Takeuchi_03}. Replacing $\dot{\varphi}$
in equation \ref{eq_rho} with $\Omega_K$ and using
 \eq{
 {GM_*\over R^2} - {GM_*\,\, R\over ( R^2+z^2)^{3/2}} \sim
 {GM_*\over R^2}\,\,\frac{3}{2}\,\,\frac{z^2}{ R^2}\ll \beta_ R{GM_*\over
  R^2+z^2}
 }
yields the solution
 \eq{\label{rhodot}
  \dot{ R} = {GM_*\over\mu}\,\,{\beta_ R\over
  R^2+z^2}.
 }
Similarly, from equation \ref{eq_z} we derive
 \eq{\label{zdot}
  \dot{z} = {GM_*\over\mu}\,\,{1\over  R^2+z^2}\,\,
  \left(\beta_z - {z\over \sqrt{ R^2+z^2}} \right).
 }
These are velocities of big dust particles in the optically thin
disk surface under the influence of stellar and diffuse radiation
pressure.

Direction of trajectories in the $ R-z$ plane is equal to the ratio
 \eq{\label{rhodot_over_zdot}
  \frac{\dot{z}}{\dot{ R}} = {\beta_z - \sin\theta \over \beta_ R},
 }
where $\sin\theta=z/\sqrt{ R^2+z^2}$. From equations \ref{force_r}
and \ref{force_p} it follows
 \eq{
  \frac{\beta_z}{\beta_ R}={ \mathfrak{F}_r\sin\theta+\mathfrak{F}_\bot\cos\theta
  \over \mathfrak{F}_r\cos\theta-\mathfrak{F}_\bot\sin\theta} =
  {\sin\theta\cos\alpha + \cos\theta\sin\alpha \over \cos\theta\cos\alpha - \sin\theta\sin\alpha}.
 }
In the inner disk region
$\cos\theta\cos\alpha\gg\sin\theta\sin\alpha$, which yields
 \eq{
 \frac{\dot{z}}{\dot{ R}} \sim \tan\theta + \tan\alpha
 -\frac{\sin\theta}{\beta_ R}.
 }

Notice that under a strong radiation pressure force (i.e.
$\sin\theta/\beta_ R\ll 1$) $\tan\theta + \tan\alpha$ is exactly the
curvature of the disk surface for small angles $\alpha$ and $\theta$
(figure \ref{Figure1}). The same is true for the inner disk rim
where $\alpha$ is not small, but $\theta\ll 1$, and the trajectory
becomes $\dot{z}/\dot{ R}\sim\tan\alpha$.

Vertical radiation pressure $\beta_z$ reduces the influence of
gravity on big grains, which results in expansion of optically thin
disk surface. This is equivalent to a disk where gravity on a given
grain size is reduced by a factor of $1-\beta_z$, which increases
the scale height by $1/\sqrt{1-\beta_z}$. Grains with $\beta_z\ge 1$
can decouple from the dense gaseous disk if they reach heights where
gas drag does not dominate the dust dynamics. Such a vertical
expansion works only with big grains and optically thin dust. {\it
If too much dust enters this zone and makes it optically thick, the
radiation pressure decreases and the expansion subdues}.

\newpage

\section{{\Large Dust mass flowing in the disk surface}}

The outflow velocity is (see equations \ref{rhodot} and \ref{zdot}):
 \eq{\label{eq_v}
  v=\sqrt{\dot{ R}^2+\dot{z}^2}\sim
  \frac{GM_*}{\mu(z_s)}\,\,\frac{\beta}{ R^2+z^2}.
 }
We assume that $v$ is constant within the optically thin surface
that starts from the height $z_s$. For the surface populated by
grains of average radius $a$ the net dust mass flux is
\cite{Takeuchi_03}
 \eq{
 \dot{M}_{sur}=2\int\limits_{z_s}^{\infty}v( R,z)\varrho_d( R,z)\,\,2\pi
  R dz,
 }
where $\varrho_d( R,z)$ is the dust number density at $( R,z)$. The
factor of two comes from the disk having two sides. Using $\mu$ from
equation \ref{eq_mu} and gas to dust ratio
$\xi=\varrho_g/\varrho_d$, the mass flux becomes
 \eq{
 \dot{M}_{sur}=4\pi {GM_* \varrho_s a \beta \over c_s \xi}
  R \int\limits_{z_s}^{\infty} {dz\over  R^2+z^2}.
 }
The assumption of a constat sound speed $c_s\sim 2000 m/s$ is
correct within about $\pm 700m/s$ for temperatures considered here.
Solving the integral and neglecting $z_s/ R\ll \pi/2$ yields
 \eq{
 \dot{M}_{sur}=\frac{0.021}{\xi}\,\, \frac{M_\oplus}{\rm year}\,\,
 \left[\frac{M_*}{M_\odot}\right]
 \left[\frac{\varrho_s}{3000kg/m^3}\right]
 \left[\frac{a}{\mu m}\right]\,\, \beta.
 }
We   replace $\beta$ with equation \ref{eq_beta} and get
 \eq{\label{eq_Msur}
   \dot{M}_{sur} = {8\times 10^{-3}\over \xi}\,\,\frac{M_\oplus}{\rm year}\,\,
   \left[\frac{L_*}{L_\odot} \right].
 }
where we assume $\int Q^{ext}_\lambda f_{*\lambda}d\lambda\sim 2$
and $|\overrightarrow{\mathfrak{f}}|\sim 1$.

Estimated amount of dust from equation \ref{eq_Msur} that flows
outward in the disk surface layer depends critically on the gas to
dust ratio $\xi$, which is very uncertain in the inner disk region
because of the interplay between dust sublimation, growth and
settling. For the standard $\xi=100$ the outflow transfers one Earth
mass within $\sim$13,000 years for $L_*=1L_\odot$ and $\sim$13 years
for $L_*=100L_\odot$. If the dust outflow becomes larger than the
disk accretion inflow then the dusty disk starts to erode from its
inner rim outward. According to our estimation, such an erosion
happens when the total disk accretion is $\la 2.4\times
10^{-8}[L_*/L_\odot]M_\odot/{\rm yr}$. This value is about the same
as the observed accretion rate averages in T Tauri \cite{Calvet04}
and Herbig Ae \cite{Garcia06} stars, while in more luminous Herbig
Be stars the limit becomes very high. Interestingly, observations
indicate structural differences in the inner disk geometry between
low and high luminosity young stellar objects
\cite{Millan-Gabet_07,Vinkovic_07}.

The timescale for an outflow of dust from cylindrical radius $R$ to
$R'>R$ can be derived from equation \ref{eq_v}. The radial component
of the velocity is
 \eq{
  v_R\sim \frac{GM_*\beta}{\mu(z_s)}\,\,\frac{R}{(R^2+z^2)^{3/2}}\sim
  \frac{GM_*\beta}{\mu(z_s)}\,\,\frac{1}{R^2},
 }
which gives the time for a grain transport from $R$ to $R'>R$
 \eq{
  t_{rad}(R,R')=\int\limits_R^{R'}dR/v_R=
   \frac{\mu(z_s) R^3}{3GM_*\beta}\left[\left({R'^3\over
   R}\right)^3-1\right]=
   \frac{\mu(z_s)}{3\beta\Omega_K^2(R)}\left[\left({R'^3\over
   R}\right)^3-1\right],
 }
where $\mu(z)$ is defined in equation \ref{eq_mu}, $z_s$ is the
height where optically thin surface starts and $\Omega_K(R)$ is the
Keplerian speed at $R$. In comparison, the local dust settling time
is $t_{\rm set}( R)=\mu/\Omega_K^2(R)$ \cite{Alex08}, where $\mu$
goes over all $z$, yielding $t_{\rm set}\sim 10^5{\rm yr}$ for $\mu
m$ sized particles in typical disks. Radiation pressure is active in
the region $z\sim z_s$ where $\mu$ is much smaller than in the disk
interior of small $z$. Therefore, it operates at much smaller
timescales than $t_{\rm set}$ when $ R'\sim R$, but becomes
comparable $t_{rad}(R,R')\sim t_{\rm set}$ when \hbox{$ R'/ R\ga
10$}.

\newpage

\section{{\Large Extent of the diffuse radiation pressure}}

We compare values of $\beta$ perpendicular to the disk surface
originating from diffuse ($\beta_{IR}$) and stellar radiation
($\beta^*\sin\alpha$)
 \eq{\label{eq_b_over_b}
  B={\beta_{IR}\over \beta^*\sin\alpha} = {\int Q_\lambda^\prime
  F_\lambda^d d\lambda \over \int Q_\lambda^\prime
  F_\lambda^*d\lambda\sin\alpha},
 }
where $Q_\lambda^\prime$ is the extinction coefficient of a dust
grain pushed by the radiation pressure. For a given distance $r$
from the star, the stellar component can be rewritten as
 \eq{\label{eq_stellar}
 \int Q_\lambda^\prime F_\lambda^*d\lambda = \left(\frac{R^*}{r}\right)^2
 \sigma_{SB}T^{*4} \langle Q^\prime\rangle_{T^*},
 }
where $\sigma_{SB}$ is the Stefan-Boltzmann constant, $R^*$ and
$T^*$ are the stellar radius and temperature, respectively, and
$\langle Q^\prime\rangle_{T^*}$ is averaged $Q_\lambda^\prime$ over
the stellar spectrum.

The diffuse radiation is difficult to specify because it depends on
structural properties of a particular protoplanetary disk, hence it
suffers from various modeling uncertainties. But we know that the
most inner dusty disk region contains only big grains in its hot
surface layer because small grains overheat at these distances and
sublimate under direct stellar radiation. Hence, here we consider a
disk surface populated with big grains. The radial thickness of disk
surface is defined by complete absorption of stellar radiation,
where we use the surface radial optical depth at
$\lambda$=0.55$\mu$m as $\tau_V\sim 1$. The diffuse radiation is
approximated according to the approach described in \cite{Chiang_97}
 \eq{\label{eq_diffuse}
 \int Q_\lambda^\prime F_\lambda^d d\lambda = \sigma_{SB}T^4 \int Q_\lambda^\prime
 b_\lambda(T)\,\, \epsilon_{IR}\sin\alpha \,\, d\lambda,
 }
where $T$ is the surface dust temperature, $b_\lambda(T)$ is
normalized Planck function and $\epsilon_{IR}\sin\alpha$ is the
surface thickness in infrared. Unlike small grains where
$\epsilon_{IR}=Q^{abs}_\lambda/Q^{abs}_V<1$ always holds, big grains
have
 \eq{
  \epsilon_{IR}=
    \cases{  Q^{abs}_\lambda/Q^{abs}_V &  when $Q^{abs}_\lambda < Q^{abs}_V$ \cr
             1 &  when $Q^{abs}_\lambda \ge Q^{abs}_V$ \cr
    }
 }
where $Q^{abs}_\lambda$ is the absorption coefficient of surface
dust. The surface temperature is dominated by stellar heating, hence
we can estimate the temperature from the equilibrium between stellar
absorption and optically thin infrared emission\footnote{Equation
\ref{eq_LTE} holds for the optically thin surface of an optically
thick disk, while optically thin disks have a factor of $1/4$ in the
right hand side of the equation. For the source of this difference
between optically thin and thick disks see
\cite{Vinkovic_06,Dullemond_01}.}
 \eq{\label{eq_LTE}
   \left(\frac{T}{T^*}\right)^4=
   \left(\frac{R^*}{r}\right)^2{\langle Q^{abs}\rangle_{T^*} \over
    \langle Q^{abs}\rangle_{T}},
 }
where $\langle Q^{abs}\rangle_{T}$ is the Planck average at
temperature $T$. Combining \ref{eq_stellar}, \ref{eq_diffuse} and
\ref{eq_LTE} with \ref{eq_b_over_b} gives
 \eq{\label{betaIRbeta}
 B={\beta_{IR}\over \beta^*\sin\alpha} =
 {\langle Q^{abs}\rangle_{T^*} \over \langle Q^{abs}\rangle_{T}}\,\,
 { \langle Q^\prime\epsilon_{IR}\rangle_{T} \over \langle
 Q^\prime\rangle_{T^*}}.
 }

\newpage

\renewcommand{\refname}{Supplementary information references}

\end{document}